\documentclass[prb,twocolumn,showpacs,nofootinbib]{revtex4}

\usepackage{amsmath,amssymb,epsfig,epsf,psfig}

\begin{document}

\title
{\bf Non -- Fermi Liquid Behavior in Fluctuating Gap Model:\\
From Pole to Zero of the Green's function}
\author{E. Z. Kuchinskii, M. V. Sadovskii}

\affiliation
{$^1$Institute for Electrophysics, Russian Academy of Sciences,
Ekaterinburg, 620016, Russia}

\begin{abstract}
We analyze non -- Fermi liquid (NFL) behavior of fluctuating 
gap model (FGM) of pseudogap behavior in both $1D$ and $2D$. We discuss in 
detail quasiparticle renormalization ($Z$ -- factor), demonstrating a kind
of ``marginal'' Fermi liquid or Luttinger liquid behavior and 
topological stability of the ``bare'' Fermi surface (Luttinger theorem).  In 
$2D$ case we discuss effective picture of Fermi  surface ``destruction'' 
both in ``hot spots'' model of dielectric (AFM, CDW) pseudogap fluctuations, 
as well as for qualitatively different case of superconducting $d$ - wave 
fluctuations, reflecting NFL spectral density behavior and similar to that
observed in ARPES experiments on copper oxides.  
\end{abstract}

\pacs{71.10.Hf, 71.27.+a, 74.72.-h}

\maketitle


\section{Introduction}

Pseudogap formation in the electronic spectrum of underdoped copper oxides 
is especially striking anomaly of the normal state of
high temperature superconductors \cite{MS}.  
Discussions on the nature of the pseudogap state continue within two main
``scenarios'' -- that of superconducting fluctuations, leading to Cooper pair
formation above $T_c$, or that of another order parameter fluctuations, in fact 
competing with superconductivity.
 
We believe that the preferable ``scenario'' for pseudogap formation 
is most likely based on the model of strong scattering of the charge
carriers by short--ranged antiferromagnetic (AFM, SDW) spin fluctuations
\cite{MS}. In momentum representation this scattering transfers 
momenta of the order of ${\bf Q}=(\frac{\pi}{a},\frac{\pi}{a})$ 
($a$ --- lattice constant of two dimensional lattice). 
This leads to the formation of structures in the one-particle spectrum, 
which are precursors of the changes in the spectra due
to long--range AFM order (period doubling).

Within this spin--fluctuation scenario a simplified model of the pseudogap 
state was studied \cite{MS,Sch,KS} under the assumption that the scattering
by dynamic spin fluctuations can be reduced for high enough temperatures
to a static Gaussian random field (quenched disorder) of pseudogap fluctuations.
These fluctuations are defined by a characteristic scattering vector from the 
vicinity of ${\bf Q}$,  with a width determined by the inverse correlation 
length of short--range order $\kappa=\xi^{-1}$. 
Actually, a similar model (formalism) can be applied also to the case of 
pseudogaps of superconducting nature \cite{KS}.

These models originated from earlier one -- dimensional model of pseudogap
behavior \cite{MS74,MS79}, the so called fluctuating gap model (FGM), which
is exactly solvable in the asymptotic limit of large correlation lengths of
pseudogap fluctuations $\kappa=\xi^{-1}\to 0$ \cite{MS74}, and ``nearly 
exactly'' solvable case of finite $\kappa$, where we can take into account
{\em all} Feynman diagrams of perturbation series, though using an 
approximate {\em Ansatz} for higher -- order contributions \cite{MS79}. 

Non -- Fermi liquid behavior of FGM model was discussed already in the case 
of $1D$ \cite{MS74,Wonn,MS91,Kenz}, as well as in $2D$ \cite{MS,Sch,KS}. 
However, some interesting aspects of this model are still under discussion
\cite{Vol05}. Below we shall analyze different aspects of this anomalous 
behavior both in $1D$ and $2D$ versions, mainly for the case of AFM (SDW) or CDW
pseudgap fluctuations, and also, more brielfly for the case of superconducting
fluctuations, demonstrating a kind of ``marginal''
Fermi liquid behavior and qualitative picture of Fermi surface
``destruction'' and formation of ``Fermi arcs'' in $2D$, similar to
that observed in ARPES experiments on copper oxides.

\section{Possible types of Green's function 
renormalization.}

Let us start with some qualitative discussion of possible manifestations of
NFL behavior.
Green's function of interacting system of electrons is expressed via Dyson 
equation (in Matsubara representation, 
$\varepsilon_n=(2n+1)\pi T$,\ $\xi_p=v_F(p-p_F)$)
as\footnote{Despite our use of Matsubara representation, below we consider
$\varepsilon_n$ as {\em continuous} variable.}:
\begin{equation}
G(\varepsilon_n,\xi_p)=\frac{1}{i\varepsilon_n-\xi_p-\Sigma(\varepsilon_n,\xi_p)}
\label{Dyson}
\end{equation}
In the following, we shall use rather {\em unusual} definition of
renormalization (``residue'') $Z$ - factor, introducing it via \cite{Vol05}:  
\begin{equation} 
G(\varepsilon_n,\xi_p)=Z(\varepsilon_n,\xi_p)G_0(\varepsilon_n,\xi_p)=
\frac{Z(\varepsilon_n,\xi_p)}{i\varepsilon_n-\xi_p}
\label{Z_factor}
\end{equation}
or
\begin{equation}
Z(\varepsilon_n,\xi_p)=\frac{i\varepsilon_n-\xi_p}
{i\varepsilon_n-\xi_p-\Sigma(\varepsilon_n,\xi_p)}=
(i\varepsilon_n-\xi_p)G(\varepsilon_n,\xi_p)
\label{Z_fac}
\end{equation}
Note that $Z(\varepsilon_n,\xi_p)$ is in general complex and actually
determines full renormalization of free -- electron Green's function
$G_0(\varepsilon_n,\xi_p)$ due to interactions. At the same time, it is
in some sense similar to standard residue renormalization factor used in 
Fermi liquid theory. 

Let us consider possible alternatives for $Z(\varepsilon_n,\xi_p)$ behavior.

\subsection{Fermi liquid behavior.}

In normal Fermi liquid we can perform the usual expansion (close to 
Fermi level and in obvious notations), assuming the absence of any 
singularities in $\Sigma(\varepsilon_n,p)$:  
\begin{equation} 
\Sigma(\varepsilon_n,\xi_p)\approx\Sigma(0,0)+i\varepsilon_n
\left.\frac{\partial\Sigma(\varepsilon_n,\xi_p)}{\partial (i\varepsilon_n)}
\right|_0+\xi_p\left.\frac{\displaystyle\partial\Sigma(\varepsilon_n,\xi_p)}
{\partial\xi_p}\right|_0+\cdots
\label{Sigexp}
\end{equation}
In the absence of static impurity scattering $\Sigma(0,0)$ is real and 
just renormalizes the chemical potential. Then we can rewrite (\ref{Dyson})
as:
\begin{equation}
G(\varepsilon)=\frac{1}{i\varepsilon_n\left\{\displaystyle 
1-\frac{\partial\Sigma}
{\partial (i\varepsilon_n)}\right\}_0-\xi_p\left\{\displaystyle 
1+\frac{\partial\Sigma}
{\partial\xi_p}\right\}_0}
\equiv\frac{\tilde Z}{i\varepsilon_n-\tilde\xi_p}
\label{expG}
\end{equation}
where we have introduced the usual renormalized residue at the pole:
\begin{equation}
\tilde Z=\frac{1}{\displaystyle 1-\left.\frac{\partial\Sigma}
{\partial (i\varepsilon_n)}\right|_0}\ ; \qquad
\tilde Z^{-1}=1-\left.\frac{\partial\Sigma}{\partial (i\varepsilon_n)}\right|_0
\label{tZ}
\end{equation}
and spectrum of quasiparicles:
\begin{equation}
\tilde\xi_p=\tilde Z\left(1+\frac{\partial\Sigma}{\partial\xi_p}\right)_0\xi_p
\end{equation}
The usual analytic continuation to real frequencies gives now the standard 
expressions of normal Fermi liquid theory \cite{Migdal,Diagr} with real
$0<\tilde Z<1$, conserving the quasiparticle pole of the Green's function. 

In the special case of $\xi_p=0$, i.e. at the Fermi surface which is
not renormalized by interactions (according to Landau hypothesis and Luttinger
theorem), we have:
\begin{equation}
G(\varepsilon_n,\xi_p)=\frac{\tilde Z}{i\varepsilon_n}
\label{Gxi0}
\end{equation}
i.e. $\tilde Z$ just coincides with the limit of
$Z(\varepsilon_n\to 0,\xi_p=0)$ as defined by (\ref{Z_factor}), (\ref{Z_fac}),
and we have the usual pole, as $\varepsilon_n\to 0$. Similarly, for
$\varepsilon_n=0$, we have $Z(\varepsilon_n=0,\xi_p\to 0)\sim\tilde Z$.

In general this behavior is conserved not only for the case of 
$\Sigma(\varepsilon_n,\xi_p)$ possessing regular expansion at small
$\varepsilon_n$ and $\xi_p$, but also for $\Sigma(\varepsilon_n,\xi_p)\sim
Max(\varepsilon_n^{\alpha}, \xi_{p}^{\alpha})$ with any $\alpha\geq 1$.

\subsection{Impure Fermi liquid.}

In case of small concentration of random static impurities we have
$\Sigma(\varepsilon_n\to 0,\xi_p\to 0)\to const$, with $Re\Sigma(0,0)$ giving
again the shift of the chemical potential, while $Im\Sigma(0,0)\sim\gamma$,
where $\gamma$ is impurity scattering rate. For the Green's function we have:
\begin{equation}
G(\varepsilon_n,\xi_p)=\frac{\tilde Z}{i\varepsilon_n-\tilde\xi_p+
i\gamma\frac{\varepsilon_n}{|\varepsilon_n|}}
\label{Gimp}
\end{equation}
so that renormalization factor defined by (\ref{Z_fac}) is given by:
\begin{equation}
Z(\varepsilon_n,\xi_p)=\tilde Z\frac{i\varepsilon_n-\xi_p}
{i\varepsilon_n-\tilde\xi_p+i\gamma\frac{\varepsilon_n}{|\varepsilon_n|}}
\label{Zimp} 
\end{equation} 
For $\xi_p=0$ we just have:  
\begin{eqnarray} 
Z(\varepsilon_n,\xi_p=0)=\tilde Z\frac{i\varepsilon_n}
{i\varepsilon_n+i\gamma\frac{\varepsilon_n}{|\varepsilon_n|}}\sim\nonumber\\
\sim\frac{|\varepsilon_n|}{\gamma}\to 0\quad\mbox{for}\quad|\varepsilon_n|\to 0
\label{Z_i}
\end{eqnarray}
while for $|\varepsilon_n|\ll|\xi_p|$:
\begin{eqnarray} 
Z(\varepsilon_n\to 0,\xi_p)=\tilde Z\frac{\xi_p}
{\xi_p-i\gamma\frac{\varepsilon_n}{|\varepsilon_n|}}\sim\nonumber\\
\sim i\frac{\xi_p}{\gamma}sign\varepsilon_n\to 0\quad\mbox{for}\quad
\xi_p\to 0 
\label{Z_ix} 
\end{eqnarray} 
i.e. impurity scattering leads to $Z$ - factor being zero at the Fermi surface, 
just removing the usual Fermi liquid {\em pole} singularity and 
producing a finite {\em discontinuity} of the Green's function  at 
$\varepsilon_n=0$. This behavior is due to the loss of translational 
invariance of the Fermi liquid theory (momentum conservation) because of 
impurities. In fact, Green's function (\ref{Gimp}) is obtained after the 
{\em averaging} over impurity position, which formally restores translational 
invariance, leading to a kind of (trivial) non -- Fermi liquid (NFL) behavior. 
Note, that this behavior is observed for $|\varepsilon_n|, |\xi_p|\ll\gamma$, 
while in the opposite limit we obviously have finite 
$Z(\varepsilon_,\xi_p)\sim\tilde Z$.

\subsection{Superconductors, Peierls and excitonic insulators.}

Consider now the case of $s$ - wave superconductor. 
Normal Gorkov Green's function is given by:
\begin{equation}
G(\varepsilon_n,\xi_p)=\frac{i\varepsilon_n+\xi_p}{(i\varepsilon_n)^2-
\xi_p^2-|\Delta|^2}
\label{Gor}
\end{equation}
where $\Delta$ is {\em superconducting} gap. The same form normal Green's 
function takes also in excitonic or Peierls insulator, where $\Delta$
denotes appropriate {\em insulating} gap in the spectrum \cite{Diagr}. Then:
\begin{eqnarray}
Z(\varepsilon_n,\xi_p)=\frac{(i\varepsilon_n)^2-(\xi_p)^2}{(i\varepsilon_n)^2-
\xi_p^2-|\Delta|^2}\sim\nonumber\\ 
\sim\frac{Max(\varepsilon_n^2,\xi_p^2)}{|\Delta|^2}
\to 0\quad\mbox{for}\quad\varepsilon_n,\xi_p\to 0
\label{Z_Gor}
\end{eqnarray}
i.e. we have NFL behavior with {\em pole} of the Green's
function at the Fermi surface replaced by {\em zero}, due to Fermi
surface being ``closed'' by superconducting (or insulating) gap.

Again, Fermi liquid type behavior with finite $Z$ - factor is ``restored''
for $|\varepsilon_n|,|\xi_p|\gg |\Delta|$.

However, the complete description of superconducting (excitonic, Peierls)
phase is achieved only after the introduction also of anomalous Gorkov
function. Excitation spectrum on both sides of the phase transition is
determined by different Green's functions with different topological 
properties \cite{Vol05}. 

\subsection{Non -- Fermi liquid behavior due to interactions.}

Non -- Fermi liquid behavior of Green's function due to interactions may 
appear also in case of singular behavior of 
$\Sigma(\varepsilon_n,\xi_p)\to\infty$ for 
$\varepsilon_n\to 0$ and $\xi_p\to 0$, e.g. power -- like divergence\footnote{
Additional logarithmic divergence can also be present here!} of
$\Sigma(\varepsilon_n,\xi_p)\sim Max (\varepsilon_n^{-\alpha},\xi_p^{-\alpha})$ 
with $\alpha>0$. Obviously, in this case we have 
$Z(\varepsilon_n\to 0,\xi_p\to 0)\to 0$, and we again have {\em 
zero} of the Green's function at the Fermi surface. 

Another possibility is singular behavior of derivatives of self - energy
in (\ref{Sigexp}), e.g. in case of
$\Sigma(\varepsilon_n,\xi_p)\sim Max(\varepsilon_n^{\alpha},\xi_p^{\alpha})$
with $0<\alpha<1$, leading to weaker than the usual pole singularity of
Green's function at the Fermi surface.

Both types of behavior are realized within Tomonaga -- Luttinger model in
one -- dimension \cite{DL}, where asymptotic behavior of 
$G(i\varepsilon_n,\xi_p)$ in the region of small 
$\xi_p\sim\varepsilon_n$ can be expressed as:     
\begin{equation} 
G(\varepsilon_n\sim \xi_p)\sim\frac{1}{\varepsilon_n^{1-2\alpha'}}
\label{Gep1}
\end{equation}
with $\alpha'<1/2$. For $\alpha'>1/2$:
\begin{equation}
G(\varepsilon_n\sim \xi_p)\sim A+B\varepsilon_n^{2\alpha'-1}
\label{Gep2}
\end{equation}
For $3/2>\alpha'>1$:
\begin{equation}
G(\varepsilon_n\sim \xi_p)\sim A+B\varepsilon_n+C\varepsilon_n^{2\alpha'-1},
\qquad \mbox{etc.} 
\label{Gep3} 
\end{equation}
with the value of $\alpha'$ determined by the strength of interaction. 

Special case is the so called ``marginal'' Fermi liquid behavior
{\em assumed} \cite{Varma} for interpretation of electronic properties of
$CuO_2$ planes of copper oxides. This is given by:
\begin{equation}
\Sigma(\varepsilon_n,\xi_p)\sim \lambda i\varepsilon_n\ln\frac{Max(\varepsilon_n,\xi_p)}
{\omega_c}
\label{margse}
\end{equation}
where $\lambda$ is some dimensionless interaction constant, and $\omega_c$ is
characteristic cut -- off frequency. If we formally use (\ref{tZ}) at finite
$\varepsilon_n$, we obtain:
\begin{equation}
\tilde Z(\varepsilon_n,\xi_p)\sim \frac{1}
{\displaystyle 1-\lambda\ln\frac{Max(\varepsilon_n,\xi_p)}{\omega_c}}
\label{Zmarg}
\end{equation}
In this case ``residue at the pole'' of the Green's function 
($Z$-factor)
\footnote{Note that (\ref{Zmarg}), strictly speaking, can not
give correct definition of the ``residue'', as standard expression (\ref{tZ}) is
defined only at the Fermi surface itself, where (\ref{Zmarg}) just does not
exist. Thus in what follows we prefer rather unusual definition given in 
(\ref{Z_factor}).}
goes to {\em zero} at the Fermi surface itself, and again quasiparticles are 
just not defined there at all! However, everywhere outside a narrow 
(logarithmic) region close to the Fermi surface we have more or less ``usual'' 
quasiparticle contribution --- quasiparticles (close to the Fermi surface) are 
just ``marginally'' defined. At present there are no generally accepted 
microscopic models of ``marginal'' Fermi liquid behavior in two --
dimensions.
  
\section{Fluctuating gap model.}

Physical nature of FGM was extensively discussed in the literature 
\cite{MS,Sch,KS,MS74,MS79,Wonn,MS91,Kenz,Diagr}. It is based on the picture
of an electron propagating in the (static!) {\em Gaussian} random field of
(pseudogap) fluctuations, leading to scattering with characteristic
momentum transfer from close vicinity of some fixed scattering vector ${\bf Q}$.
These fluctuations are described by two basic parameters:\ amplitude
$\Delta$ and correlation length (of short -- range order) $\xi^{-1}$, 
determining effective width $\kappa=\xi^{-1}$ of scattering vector distribution.

In one -- dimension, the typical choice of scattering vector is $Q=2p_F$
(fluctuation region of Peierls transition) \cite{MS74,MS79}, while 
in two -- dimensions we usually mean the so called ``hot spots'' model with 
${\bf Q}=(\pi/a,\pi/a)$ \cite{Sch,KS}.  These models assume ``dielectric'' 
(CDW, SDW) nature of pseudogap fluctuations, but essentially the same 
formalism can be used in case of superconducting fluctuations \cite{KS}. 

The case of superconducting ($s$ - wave) pseudogap fluctuations in higher
dimensions is described actually by the same one -- dimensional version of
FGM \cite{MS74,KS,Vol05}.

Attractive property of models under discussion is the possibility of an exact
solution achieved by the complete summation of the whole Feynman diagram
series in the asymptotic limit of large correlation lengths $\xi\to\infty$
\cite{MS74,Wonn}. In case of finite correlation lengths we can also perform
summation of {\em all} Feynman diagrams for single -- electron Green's function,
using an {\em approximate} Ansatz for higher order contributions both in
one dimension \cite{MS79} and in two dimensions \cite{Sch,KS}. Similar methods
of diagram summation can be also applied in calculations of two -- particles
Green's functions (vertex parts) \cite{MS74,MS91,Sch,KS,SS,Diagr}.

Our aim will be to demonstrate that nearly all aspects of NFL
behavior discussed above can be nicely demonstrated within different variants 
of FGM. 

\subsection{One -- dimension.}

We shall limit ourselves here only to the case of incommensurate pseudogap
(CDW) fluctuations \cite{MS74,MS79}. Commensurate case \cite{Wonn,MS79} can be 
analyzed in a similar way. Note that the same expressions apply also for the
case of superconducting ($s$ - wave) fluctuations in all dimensions. 

In the limit of infinite correlation length of pseudogap fluctuations we have
the following {\em exact} solution for a single -- electron Green's function
\cite{MS74,Diagr}:
\begin{eqnarray}
G(\varepsilon_n,\xi_p)=\int_0^{\infty}d\zeta e^{-\zeta}
\frac{i\varepsilon_n+\xi_p}{(i\varepsilon_n)^2-
\xi_p^2-\zeta\Delta^2}=\nonumber\\
=\frac{i\varepsilon_n+\xi_p}{\Delta^2}\exp\left(\frac{\varepsilon_n^2+\xi_p^2}
{\Delta^2}\right)Ei\left(\frac{\varepsilon_n^2+\xi_p^2}
{\Delta^2}\right)\approx\nonumber\\
\approx \frac{i\varepsilon_n+\xi_p}{\Delta^2}\ln\left(\gamma'
\frac{\varepsilon_n^2+\xi_p^2}{\Delta^2}\right)\quad\mbox{for}\quad
\varepsilon_n\to 0,\ \xi_p\to 0
\label{Gzav}
\end{eqnarray}
where $Ei(-x)$ denotes integral exponential function and we used the
asymptotic behavior $Ei(-x)\sim \ln(\gamma' x)$ for $x\to 0$ 
($\ln\gamma'=0.577$ -- Euler constant). Then, using (\ref{Z_fac}) we immediately 
obtain:
\begin{eqnarray}
Z(\varepsilon_n\sim 
\xi_p)=-\frac{\varepsilon_n^2+\xi_p^2}{\Delta^2}\ln\left(\gamma' 
\frac{\varepsilon_n^2+\xi_p^2}{\Delta^2}\right)\to\nonumber\\
\to 0\quad\mbox{for}\quad
\varepsilon_n\to 0,\ \xi_p\to 0
\label{Z1D}
\end{eqnarray}
Precisely the same result is obtained if we define for finite $\varepsilon_n,
\xi_p$:
\begin{equation}
\tilde Z(\varepsilon_n,\xi_p)=\frac{1}{\displaystyle 
1-\frac{\partial\Sigma(\varepsilon_n,\xi_p)}{\partial(i\varepsilon_n)}} 
\label{ttZ}
\end{equation}
similar to (\ref{tZ}). 
Note that due to $|\varepsilon_n|\ll\Delta$,\ $|\xi_p|\ll\Delta$ 
we obviously have $Z>0$, but the usual {\em pole} of the Green's function 
at the Fermi surface (``point'') of the ``normal'' system is transformed 
here to {\em zero} due to pseudogap fluctuations. Because of topological 
stability \cite{Vol05}, the singularity of the Green's function at the Fermi 
surface is not destroyed:  {\em zero is also a singularity} (with the same 
topological charge) as pole. But actually FGM gives an explicit example of a 
kind of Luttinger or ``marginal'' Fermi liquid with very strong 
renormalization of singularity at the Fermi surface.

Consider self -- energy corresponding to Green's functions (\ref{Gzav}):
\begin{equation}
\Sigma(\varepsilon_n,\xi_p)=i\varepsilon_n-\xi_p-\left[
\int_0^{\infty}d\zeta e^{-\zeta}
\frac{i\varepsilon_n+\xi_p}{(i\varepsilon_n)^2-
\xi_p^2-\zeta\Delta^2}\right]^{-1}
\label{S_en}
\end{equation}
so that taking, for brevity, $\xi_p=0$ and $\varepsilon_n\to 0$ we get:
\begin{eqnarray}
&& \Sigma(\varepsilon_n\to 0,\xi_p=0)=\frac{1}{i\varepsilon_n}\left[
\int_0^{\infty}d\zeta e^{-\zeta}
\frac{1}{\varepsilon_n^2+\zeta\Delta^2}\right]^{-1}\approx\nonumber\\
&& \approx -\frac{\Delta^2}{i\varepsilon_n}\frac{1}{\displaystyle\ln
\left(\gamma'\frac{\varepsilon_n^2}{\Delta^2}\right)}\to\infty
\label{S_ens}
\end{eqnarray}
i.e. the divergence of the type discussed above. 

In case of finite correlation lengths $\xi=\kappa^{-1}$ of pseudogap 
fluctuations we have to use continuous fraction representation of single -- 
electron Green's function derived in Ref. \cite{MS79} to obtain 
renormalization factor as ($\varepsilon_n>0$):
\tiny
\begin{eqnarray}
&& Z(\varepsilon_n,\xi_p)=\nonumber\\
&& =\frac{i\varepsilon_n-\xi_p}{\displaystyle i\varepsilon_n-\xi_p-
\frac{\Delta^2}{\displaystyle i\varepsilon_n+\xi_p+iv_F\kappa-
\frac{\Delta^2}{\displaystyle i\varepsilon_n-\xi_p+
2iv_F\kappa-\frac{2\Delta^2}{\displaystyle i\varepsilon_n+
\xi_p+3iv_F\kappa-...}}}}
\nonumber\\
&&
\label{Z_chain}
\end{eqnarray}
\normalsize
which can be studied numerically.

\begin{figure}[htb]
\includegraphics[clip=true,width=0.5\textwidth]{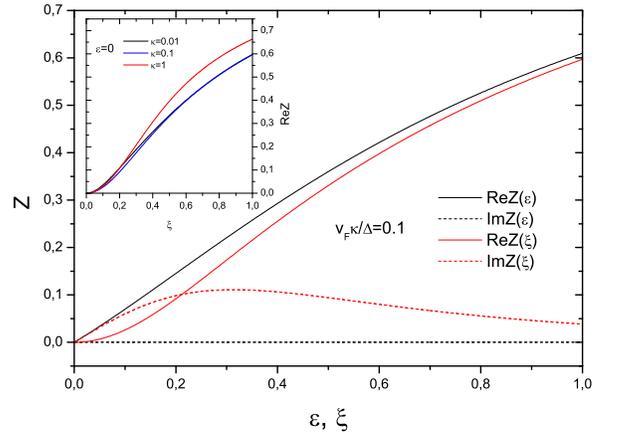}
\caption{Typical dependencies of $Z(\varepsilon_n,\xi_p)$ -- factor in 
one -- dimensional FGM with finite correlation lengths:
dependences of $Z(\varepsilon_n=0,\xi_p)$ and
$Z(\varepsilon_n,\xi_p=0)$ on $\varepsilon_n$ and $\xi_p$ for 
$v_F\kappa/\Delta=0.1$.\ At the insert:
dependences of $ReZ(\varepsilon_n=0,\xi_p)$ on $\xi_p$ for 
different values of $\kappa$ (in units of $\frac{\Delta}{v_F}$). 
Both $\varepsilon_n$ and $\xi_p$ are given 
in units of $\Delta$.} 
\label{Z_fac_1d} 
\end{figure}

In Fig. \ref{Z_fac_1d} we show typical dependences of renormalization factor
$Z(\varepsilon_n,\xi_p)$. In all cases it goes to zero at the (``bare'') Fermi 
surface and pole of Green's function disappears. Essentially, this strong 
renormalization starts on the scale of pseudogap width, 
i.e. for $|\varepsilon_n|<\Delta$ and $|\xi_p|<\Delta$, reflecting 
non -- Fermi liquid behavior due to pseudogap fluctuations. 

\begin{figure}[htb]
\includegraphics[clip=true,width=0.5\textwidth]{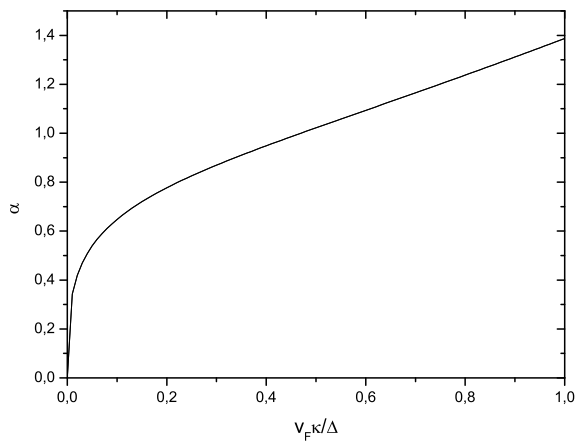}
\caption{Dependence of $\alpha\left(\frac{v_F\kappa}{\Delta}\right)$ on
inverse correlation length.}
\label{alph}
\end{figure}

However, the role of finite correlation lengths $\xi$ (finite
$\kappa$) is qualitatively similar to static impurity scattering\footnote{This
is due to our approximation of the static nature of pseudogap fluctuations.} and
more detailed calculation shows, that the behavior of $Z$ - factor
at small $\epsilon_n\ll v_F\kappa$ and $|\xi_p|\ll v_F\kappa$ (with 
$\varepsilon_n>0$) is as follows:
\begin{eqnarray}
Z(\epsilon_n,\xi_p)\approx \alpha\left(\frac{v_F\kappa}{\Delta}\right)
\left(\frac{\epsilon_n+i\xi_p}{\Delta}\right)\to\nonumber\\
\to 0\quad\mbox{for}\quad
\varepsilon_n\to 0,\ \xi_p\to 0,
\label{Z_alf}
\end{eqnarray}
with $\alpha(v_F\kappa/\Delta)\to 0$ for $\kappa\to 0$, as seen from 
Fig. \ref{alph}. In terms of Green's function this behavior corresponds to:
\begin{equation}
G(\varepsilon_n,\xi_p)\approx\frac{1}{\Delta}\alpha\left(\frac{v_F\kappa}
{\Delta}\right)\frac{\varepsilon_n+i\xi_p}{i\varepsilon_n
-\xi_p}=-i\frac{1}{\Delta}\alpha\left(\frac{v_F\kappa}{\Delta}\right)
\label{Galph}
\end{equation}
Thus, for finite $\kappa$, there is no zero of Green's function for 
$\epsilon_n=0$ and $\xi_p=0$, it remains finite as in impure system.

Vanishing of renormalization factor $Z(\varepsilon_n,\xi_p)$ at the ``bare''
Fermi surface is in correspondence with general topological stability arguments
\cite{Vol05} -- in the absence of static impurity -- like scattering the pole
singularity of the Green's function is replaced by zero. In the presence of
this additional scattering this zero is replaced by finite discontinuity,
i.e. singularity still persists.

\subsection{``Hot spots'' model in two -- dimensions.}

\begin{figure}[htb]
\includegraphics[clip=true,width=0.3\textwidth]{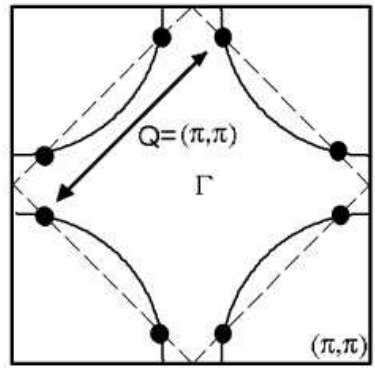}
\caption{Fermi surface in the Brillouin zone and ``hot spots'' model. 
Magnetic zone appears e.g. in the presence of antiferromagnetic long --
range order. ``Hot spots'' correspond to intersections of its borders
with Fermi surface and are connected by the scattering vector of the order
of ${\bf Q}=(\frac{\pi}{a},\frac{\pi}{a})$.}
\label{h_sp}
\end{figure}

In two dimensions we introduce the so called ``hot spots'' model.
Consider typical Fermi surface of electrons moving in the $CuO_2$ plane of
copper oxides as shown in Fig. \ref{h_sp}. 
If we neglect fine details, the observed 
(e.g. in ARPES experiments) Fermi surface (and also the spectrum of elementary
excitations) in $CuO_2$ plane, in the first approximation are 
described by the usual tight -- binding model:
\begin{equation}
\epsilon({\bf p})=-2t(\cos p_xa+\cos p_ya)-4t'\cos p_xa\cos p_ya
\label{spectr}
\end{equation}
where $t$ is the nearest neighbor transfer integral, while
$t'$ is the transfer integral between second -- nearest neighbors, 
$a$ is the square lattice constant.

Phase transition to antiferromagnetic (SDW) state
induces lattice period doubling and leads to the appearance of 
``antiferromagnetic'' Brillouin zone in inverse
space as is also shown in Fig. \ref{h_sp}. If the spectrum of carriers is 
given by (\ref{spectr}) with $t'=0$ and we consider the half -- filled case, 
Fermi surface becomes just a square coinciding with the borders of 
antiferromagnetic zone and we have a complete ``nesting'' --- flat parts of 
the Fermi surface match each other after the translation by vector of 
antiferromagnetic ordering ${\bf Q}=(\pm\pi/a,\pm\pi/a)$. In this case and 
for $T=0$ the electronic spectrum is unstable, energy gap appears everywhere 
on the Fermi surface and the system becomes insulator, due to the formation 
of antiferromagnetic spin density wave (SDW)\footnote{Analogous 
dielectrization is realized also in the case of the formation of the similar 
charge density wave (CDW).}. In the case of the Fermi surface shown in 
Fig.\ref{h_sp} the appearance of antiferromagnetic long - range order, in 
accordance with general rules of the band theory, leads to the appearance of 
discontinuities of isoenergetic surfaces (e.g. Fermi surface) at crossing 
points with borders of new (magnetic) Brillouin zone due to gap opening at 
points connected by vector ${\bf Q}$. 

In the most part of underdoped region of cuprate phase diagram  
antiferromagnetic long -- range order is absent, however, a number of 
experiments support the existence of well developed fluctuations of 
antiferromagnetic short -- range order which scatter electrons with 
characteristic momentum transfer of the order of ${\bf Q}$. 
Similar effects may appear due CDW fluctuations. These pseudogap fluctuations
are again considered to be static and Gaussian, and characterized by two
parameters: amplitude $\Delta$ and correlation length $\xi=\kappa^{-1}$
\cite{MS}. In this case we can obtain rather complete solution for single --
electron Green's function via summation of {\em all} Feynman diagrams of
perturbation series, describing scattering by these fluctuations \cite{MS,
Sch,KS}. This solution is again exact in the limit of $\xi\to\infty$ 
\cite{Sch}, and apparently very close to an exact one in case of finite 
$\xi$ \cite{MS00}. Generalizations of this approach for two -- particle
properties (vertex -- parts) are also quite feasible.

We shall start again with an exact solution for $\xi\to\infty$ (or $\kappa=0$)
\cite{Sch}. First, let us introduce (normal) Green's function for SDW (CDW)
state with long -- range order (see e.g. \cite{Diagr}):
\begin{equation}
G(\varepsilon_n,\xi_{\bf p})=\frac{i\varepsilon_n-\xi_{\bf p-Q}}
{(i\varepsilon_n-\xi_{\bf p})(i\varepsilon_n-\xi_{\bf p-Q})-W^2}
\label{GLRO}
\end{equation}
where $W$ denotes the amplitude of SDW (CDW) periodic potential and
$\xi_{\bf p}=\varepsilon({\bf p})-\mu$. 
Then we can write down appropriate $Z$ - factor as:
\begin{equation}
Z(\varepsilon_n,\xi_{\bf p})=\frac{(i\varepsilon_n-\xi_1)
(i\varepsilon_n-\xi_2)}
{(i\varepsilon_n-\xi_1)(i\varepsilon_n-\xi_2)-W^2}
\label{ZSDW}
\end{equation}
where we have denoted for brevity: $\xi_{\bf p}=\xi_1$ and 
$\xi_{\bf p-Q}=\xi_2$. In the following we shall be mainly interested in the
limit of $\varepsilon_n\to 0$ and $\xi_1\to 0$, i.e. on the approach to the
``bare'' Fermi surface. Note that $\xi_2=0$ defines the so called ``shadow''
Fermi surface. We have $\xi_1=\xi_2=0$ precisely at the ``hot spots''. 
In the following it is convenient to introduce a complex variable:
\begin{equation}
z=(i\varepsilon_n-\xi_1)(i\varepsilon_n-\xi_2)
\label{z_var}
\end{equation}
which becomes small for $\varepsilon_n,\xi_1,\xi_2\to 0$.

\subsubsection{Incommensurate combinatorics.}

In case of incommensurate (CDW) pseudogap fluctuations, an exact solution for
the Green's function of FGM in the limit of correlation length $\xi\to\infty$
takes the form similar to (\ref{Gzav}) \cite{MS,Sch} and we get
(averaging (\ref{ZSDW}) with Rayleigh distribution for W):
\begin{eqnarray}
Z(z)=\int_0^{\infty}dW\frac{2W}{\Delta^2}e^{-\frac{W^2}{\Delta^2}}
\frac{z}{z-W^2}=\nonumber\\
=\int_0^{\infty}\frac{d\zeta}{\Delta^2}e^{-\frac{\zeta}
{\Delta^2}}\frac{z}{z-\zeta}=\frac{z}{\Delta^2}e^{-\frac{z}{\Delta^2}}
Ei\left(\frac{z}{\Delta^2}\right)
\label{ZCDWav}
\end{eqnarray}
Then, for $z\to 0$ we get:
\begin{equation}
Z(z\to 
0)\approx\frac{z}{\Delta^2}\left[\ln\left(\gamma'\frac{z}{\Delta^2}\right)
-i\pi\right] 
\label{Zz}
\end{equation}
At the ``bare'' Fermi surface we have $\xi_1=0$ and in the following we limit 
ourselves to $\varepsilon_n>0$. Then, from (\ref{Zz}) we can easily find 
limiting behavior of $Z(z)$. Just quoting some results we have:
\begin{enumerate}
\item{For $\varepsilon_n\ll|\xi_2|$:
\begin{equation}
ReZ(\varepsilon_n\ll|\xi_2|,\xi_1=0)\approx\frac{\pi}{2}
\frac{\varepsilon_n|\xi_2|}{\Delta^2}
\label{ReZe}
\end{equation}
i.e. ``impure'' -- like linear behavior in $\varepsilon_n$.
}
\item{For $\varepsilon_n\gg|\xi_2|$ (i.e. also at the ``hot spot'', where
$\xi_2=0$):
\begin{equation}
ReZ(\varepsilon_n\gg|\xi_2|,\xi_1=0)\approx -\frac{\varepsilon_n^2}{\Delta^2}
\ln\left(\gamma'\frac{\varepsilon_n^2}{\Delta^2}\right)+\frac{1}{2}
\frac{\xi_2^2}{\Delta^2}
\label{ReZx}
\end{equation}
i.e. (for $\xi_2=0$) NFL behavior similar to one -- dimensional case.
}
\end{enumerate}
Note that we always have $Im Z=0$ for $\xi_2=0$, i.e. at the ``shadow'' 
Fermi surface and in particular at the ``hot spot'' itself.

\subsubsection{Spin -- fermion combinatorics.}

Consider now spin -- fermion (Heisenberg) model for pseudogap (SDW)
fluctuations \cite{Sch}. In this case we again obtain FGM, but with
gap distribution is different (from Rayleigh distribution) and instead of
(\ref{ZCDWav}) we have:
\begin{eqnarray}
&& Z(z)=\frac{2}{\sqrt{2\pi}}\int_0^{\infty}dW
\frac{W^2}{\left(\frac{\Delta^2}{3}\right)^{3/2}}
e^{-\frac{W^2}{2\left(\frac{\Delta^2}{3}\right)}}\frac{z}{z-W^2}=\nonumber\\
&& =\frac{1}{\sqrt{2\pi}}\int_{0}^{\infty}d\zeta
\frac{\sqrt{\zeta}}{\left(\frac{\Delta^2}{3}\right)^{3/2}}
e^{-\frac{\zeta}{2\left(\frac{\Delta^2}{3}\right)}}
\frac{\zeta}{z-\zeta}=
\nonumber\\
&& =\frac{\Gamma(3/2)}{\sqrt{2\pi}}\frac{(-z)^{3/2}}
{\left(\frac{\Delta^2}{3}\right)^{3/2}}\exp\left[-\frac{z}{2\left(
\frac{\Delta^2}{3}\right)}\right]
\Gamma\left(-\frac{1}{2};-\frac{z}{2\left(\frac{\Delta^2}{3}\right)}\right)
\nonumber
\label{ZSDWav}
\end{eqnarray}
Thus, for $z\to 0$ we obtain:
\begin{equation}
Z(z)\approx \frac{2\Gamma(3/2)}{\sqrt{\pi}}\left[-\frac{z}
{2\left(\frac{\Delta^2}{3}\right)}+\Gamma(-1/2)\left(
-\frac{z}{2\left(\frac{\Delta^2}{3}\right)}\right)^{3/2}\right]
\label{ZSDWz}
\end{equation}
Then on ``bare'' Fermi surface ($\xi_p=0$) we have:
\begin{eqnarray}
&& Z(\varepsilon_n\to 0,\xi_2,\xi_1=0)=\nonumber\\
&& =\frac{2\Gamma(3/2)}{\sqrt{\pi}}
\left[-\frac{\varepsilon_n(\varepsilon_n+i\xi_2)}
{2\left(\frac{\Delta^2}{3}\right)}+\right.\nonumber\\
&& \left.+\Gamma(-1/2)\left(
-\frac{\varepsilon_n(\varepsilon_n+i\xi_2)}
{2\left(\frac{\Delta^2}{3}\right)}\right)^{3/2}\right]
\label{ZSDWfs}
\end{eqnarray}
In particular, for $\xi_2=0$ we have $ImZ=0$ and:
\begin{eqnarray}
&& Z(\varepsilon_n\to 0,\xi_2=\xi_1=0)=Re Z(\varepsilon_n\to 0,\xi_2=\xi_1=0)=
\nonumber\\
&& =\frac{\Gamma(3/2)}{\sqrt{\pi}}\frac{\varepsilon_n^2}
{\left(\frac{\Delta^2}{3}\right)}
\label{ReZSDWfs}
\end{eqnarray}
so that we obtain quadratic NFL behavior of $Z$ - factor.
Again let us present some results on limiting behavior:
\begin{enumerate}
\item{For $\varepsilon_n\ll|\xi_2|$:
\begin{eqnarray}
&& ReZ(\varepsilon_n\ll|\xi_2|,\xi_1=0)=\nonumber\\
&& =\frac{2\Gamma(3/2)}{\sqrt{\pi}}\left[
\frac{\varepsilon_n^2}{2\left(\frac{\Delta^2}{3}\right)}+\sqrt{2\pi}
\left(\frac{\varepsilon_n|\xi_2|}{2\left(\frac{\Delta^2}{3}\right)}
\right)^{3/2}\right]
\label{ReSDWZe}
\end{eqnarray}
i.e. NFL ``zero'' behavior.
}
\item{For $\varepsilon_n\gg|\xi_2|$ (i.e. also at the ``hot spot'' where
$\xi_2=0$):
\begin{equation}
ReZ(\varepsilon_n\gg\xi_2,\xi_1=0)=\frac{\Gamma(3/2)}{\sqrt{\pi}}
\frac{\varepsilon_n^2}{\left(\frac{\Delta^2}{3}\right)}
\label{ReSDWZx}
\end{equation}
i.e. again NFL ``zero'' behavior.
}
\end{enumerate}

In the general case of finite correlation lengths 
$\xi=\kappa^{-1}$ we have to perform numerical analysis using the recursion 
relations proposed in Refs. \cite{Sch,KS}. 
Again we use the basic definition of $Z$ - factor given in (\ref{Z_fac}).
To calculate self -- energy $\Sigma(\varepsilon_n,\xi_{\bf p})$ 
of an electron moving in the quenched
random field of (static) Gaussian spin fluctuations with dominant
scattering momentum transfers from the vicinity of characteristic
vector ${\bf Q}$, we use the following recursion procedure 
\cite{Sch,KS} which takes into account {\em all} Feynman 
diagrams describing the scattering of electrons by this random field. 
The desired self--energy is given by
\begin{equation}
\Sigma(\varepsilon_n,\xi_{\bf p})=\Sigma_{k=1}(\varepsilon_n,\xi_{\bf p})
\label{Sk}
\end{equation}
with $\xi_{\bf p}=\epsilon({\bf p})-\mu$ (cf. (\ref{spectr})) and
\begin{equation}
\Sigma_{k}(\varepsilon_n,\xi_{\bf p})=\Delta^2\frac{s(k)}
{i\varepsilon_n+\mu
-\epsilon_k({\bf p})+inv_k\kappa-\Sigma_{k+1}(\varepsilon_n,\xi_{\bf p})}\;\;. 
\label{rec}
\end{equation} 
The quantity $\Delta$ again characterizes the energy scale of pseudogap
fluctuations and $\kappa=\xi^{-1}$ is the inverse correlation length of 
short range SDW fluctuations, $\epsilon_k({\bf p})=\epsilon({\bf 
p+Q})$ and $v_k=|v_{\bf p+Q}^{x}|+|v_{\bf p+Q}^{y}|$ for odd $k$ while 
$\varepsilon_k({\bf p})=\varepsilon({\bf p})$ and $v_{k}= |v_{\bf p}^x|+
|v_{\bf p}^{y}|$ for even $k$. The velocity projections $v_{\bf p}^{x}$ 
and $v_{\bf p}^{y}$ are determined by usual momentum derivatives of the 
``bare'' electronic energy dispersion $\epsilon({\bf p})$ given by 
(\ref{spectr}). Finally, $s(k)$ 
represents a combinatorial factor with 
\begin{equation} 
s(k)=k 
\label{vcomm} 
\end{equation}
for the case of commensurate charge (CDW type) fluctuations with
${\bf Q}=(\pi/a,\pi/a)$ \cite{MS79}. 
For incommensurate CDW fluctuations \cite{MS79} one finds
\begin{equation} 
s(k)=\left\{\begin{array}{cc}
\frac{k+1}{2} & \mbox{for odd $k$} \\
\frac{k}{2} & \mbox{for even $k$}.
\end{array} \right.
\label{vinc}
\end{equation}
For spin -- fermion model of Ref.~\cite{Sch},
the combinatorics of diagrams becomes more complicated.
Spin - conserving scattering processes obey commensurate combinatorics,
while spin - flip scattering is described by diagrams of incommensurate
type (``charged'' random field in terms of Ref.~\cite{Sch}). In this model
the recursion relation for the single-particle Green function is again given 
by (\ref{rec}), but the combinatorial factor $s(n)$ now acquires the 
following form \cite{Sch}:
\begin{equation} 
s(k)=\left\{\begin{array}{cc}
\frac{k+2}{3} & \mbox{for odd $k$} \\
\frac{k}{3} & \mbox{for even $k$}.
\end{array} \right.
\label{vspin}
\end{equation}
Below we only present our results for spin -- fermion combinatorics, as in
other cases we obtain more or less similar behavior of renormalization factors.

\begin{figure}[htb]
\includegraphics[clip=true,width=0.5\textwidth]{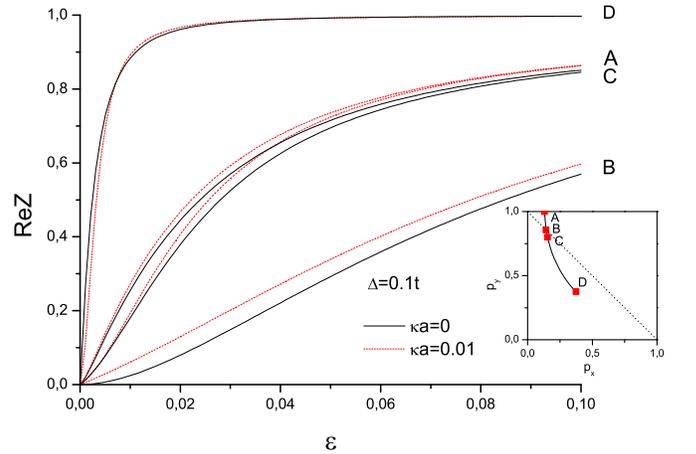}
\caption{Dependence of $Re Z$ on $\varepsilon_n$ (in units of $t$) at 
different points Fermi surface (corresponding to $t'=-0.4t$ and $\mu=-1.3t$) 
in ``hot spots'' model (spin -- fermion combinatorics of diagrams)
with correlation lengths $\xi\to\infty$ $(\kappa=0)$ 
and $\xi^{-1}a=\kappa a=0.01$. Pseudogap amplitude $\Delta=0.1t$. 
At the insert we show the 
``bare'' Fermi surface and points, where calculations were done.}
\label{Ze_comp}
\end{figure}

In Fig. \ref{Ze_comp} we show the results of numerical calculation of
$Re Z(\varepsilon_n, \xi_{\bf p}=0)$ at different points taken at the 
``bare'' Fermi surface, shown at the insert. For comparison we show data
obtained in the limit of infinite correlation length $\xi\to\infty$ (or
$\kappa=0$ -- exactly solvable case) and for finite $\kappa a = 0.01$
(i.e. $\xi=100 a$). It is clearly seen that in both cases
$Re Z\sim 1$ at ``nodal'' point $D$, except at very small values of 
$\varepsilon_n$, while in the vicinity of the ``hot spot'' (points $A$ and
$C$), and also  at the ``hot spot'' itself (point $B$), $Re Z$ becomes small
in rather wide interval of $\varepsilon_n < \Delta$. This corresponds to more 
or less ``Fermi liquid'' behavior for ``nodal'' region (vicinity of 
Brillouin zone diagonal), with a kind of ``marginal'' Fermi liquid or
Luttinger liquid (NFL) behavior as we move to the vicinity of the ``hot spot''.     

\begin{figure}[htb]
\includegraphics[clip=true,width=0.5\textwidth]{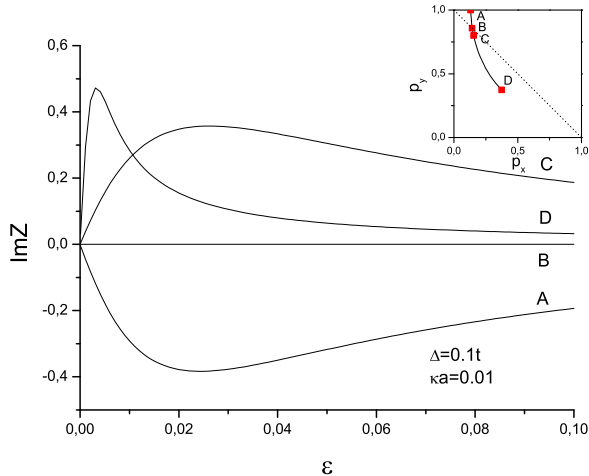}
\caption{Dependence of $Im Z$ on $\varepsilon_n$ (in units of transfer 
integral $t$) at different points Fermi surface (corresponding to $t'=-0.4t$ 
and $\mu=-1.3t$) in ``hot spots'' model with finite correlation length 
$\xi^{-1}a=\kappa a= 0.01$ (spin -- fermion combinatorics of diagrams).  
Pseudogap amplitude $\Delta=0.1t$. At the insert --- the 
``bare'' Fermi surface and points, where calculations were done. 
}
\label{ImZe}
\end{figure}

For completeness in Fig. \ref{ImZe} we show similar comparison of dependences 
of $Im Z$ on $\varepsilon_n$ at the same characteristic points on the 
Fermi surface and for the same parameters as in Figs. \ref{Ze_comp}.
It is only important to stress once again, that only at the ``hot spot''
itself (point $B$) we have $Im Z=0$, so that $Z$ becomes real, and shows 
dependence on $\varepsilon_n$ more or less equivalent to that proposed for 
``marginal'' Fermi liquids (or Luttinger liquids).

In all cases we observe vanishing of renormalization factor 
$Z(\varepsilon_n,\xi_p)$ at the ``bare'' Fermi surface.  
In the absence of static impurity -- like scattering due to finite values of
correlation length $\xi=\kappa^{-1}$ the pole singularity of the Green's 
function is replaced by zero, reflecting topological stability of the
``bare'' Fermi surface (Luttinger theorem) \cite{Vol05}. In the presence of
this scattering, singularity of the Green's function at topologically stable
``bare'' Fermi surface remains in the form of finite discontinuity.

\subsection{Spectral density and Fermi surface ``destruction'' in
``hot spots'' model.}

Let us return to (\ref{GLRO}) and perform the usual analytic continuation to
real frequencies: $i\varepsilon_n\to\varepsilon+i\delta$. Then we obtain:
\begin{eqnarray}
G^R(\varepsilon,\xi_{\bf p})=\frac{\varepsilon-\xi_2}
{(\varepsilon+i\delta-\xi_1)(\varepsilon-\xi_2+i\delta)-W^2}=\nonumber\\
=\frac{\varepsilon-\xi_2}
{(\varepsilon-\xi_1)(\varepsilon-\xi_2)-W^2+i\delta(2\varepsilon-\xi_1-\xi_2)}
\label{GLROR}
\end{eqnarray}
so that spectral density in the case of long -- range (CDW,SDW) order has the
following form:
\begin{eqnarray}
A_W(\varepsilon,\xi_{\bf p})=-\frac{1}{\pi}Im G^R(\varepsilon,\xi_{\bf p})=
\nonumber\\
=(\varepsilon-\xi_2)\delta[(\varepsilon-\xi_1)(\varepsilon-\xi_2)-W^2]
sign(2\varepsilon-\xi_1-\xi_2) \nonumber\\
\label{SpDenW}
\end{eqnarray}
Accordingly, for FGM with correlation length $\xi\to\infty$ we have:
\begin{equation}
A(\varepsilon,\xi_{\bf p})=\int_{0}^{\infty}dW{\cal P}_W
A_W(\varepsilon,\xi_{\bf p})
\label{SpDen}
\end{equation}
where ${\cal P}_W$ is distribution function of gap fluctuations, depending
on combinatorics of diagrams and leading to the following separate cases, 
already considered (or mentioned) above:

\subsubsection{Incommensurate combinatorics.}

In the case of incommensurate CDW -- like pseudogap fluctuations we have:
\begin{equation}
{\cal P}_W=\frac{2W}{\Delta^2}e^{-\frac{W^2}{\Delta^2}}
\label{Rayl}
\end{equation}
-- Rayleigh distribution \cite{MS74,Diagr}. Then, from (\ref{SpDen}) we 
obtain:
\begin{eqnarray}
&&A(\varepsilon,\xi_{\bf p})=\nonumber\\
&&=\frac{\varepsilon-\xi_2}{\Delta^2}
e^{-\frac{(\varepsilon-\xi_1)(\varepsilon-\xi_2)}{\Delta^2}}\times\nonumber\\
&&\times\theta[(\varepsilon-\xi_1)(\varepsilon-\xi_2)]sign(2\varepsilon-\xi_1-\xi_2)
\nonumber\\
\label{SpDenInc}
\end{eqnarray}
For $\varepsilon=0$ we have:
\begin{equation}
A(\varepsilon=0,\xi_{\bf p})=\frac{\xi_2}{\Delta^2}
e^{-\frac{\xi_1\xi_2}{\Delta^2}}\theta[\xi_1\xi_2]sign(\xi_1+\xi_2)
\label{SpDenInc0}
\end{equation}
For $\xi_1\to\pm 0$ we get:
\begin{equation}
A(\varepsilon=0,\xi_{\bf p}\to \pm0,\xi_2)=\pm\frac{\xi_2}{\Delta^2}
\theta(\pm\xi_2)
\label{SpDenInc0+}
\end{equation}
so that within the Brillouin zone $A(\varepsilon=0,\xi_{\bf p})$ is nonzero
only in the space between ``bare''  Fermi surface and ``shadow'' Fermi surface. This
qualitative result is confirmed below, for all other combinatorics, for the
case of ``pure'' FGM with $\xi^{-1}=\kappa=0$.

\subsubsection{Commensurate combinatorics.} 

In the case of commensurate CDW -- like pseudogap fluctuations we have
\cite{Wonn}:
\begin{equation}
{\cal P}_W=\frac{1}{\sqrt{2\pi}\Delta}e^{-\frac{W^2}{2\Delta^2}}
\label{Gaus}
\end{equation}
-- Gaussian distribution. Then, from (\ref{SpDen}) we obtain:
\begin{eqnarray}
&&A(\varepsilon,\xi_{\bf p})=\nonumber\\
&&=\frac{1}{\sqrt{2\pi}}\frac{\varepsilon-\xi_2}
{\Delta\sqrt{(\varepsilon-\xi_1)(\varepsilon-\xi_2)}}
e^{-\frac{(\varepsilon-\xi_1)(\varepsilon-\xi_2)}{2\Delta^2}}\times\nonumber\\
&&\times\theta[(\varepsilon-\xi_1)(\varepsilon-\xi_2)]sign(2\varepsilon-\xi_1-\xi_2)
\label{SpDenCom}
\end{eqnarray}
with the same qualitative conclusions as in incommensurate case.

\subsubsection{Spin -- fermion combinatorics.}

In the case of SDW -- like pseudogap fluctuations of (Heisenberg)
spin -- fermion model \cite{Sch} we have gap distribution:
\begin{equation}
{\cal P}_W=\frac{2}{\pi}\frac{W^2}{\left(\frac{\Delta^2}{3}\right)^{3/2}}
e^{-\frac{W^2}{2\left(\frac{\Delta^2}{3}\right)}} 
\label{SFGaus}
\end{equation}
Then, from (\ref{SpDen}) we obtain:
\begin{eqnarray}
&&A(\varepsilon,\xi_{\bf p})=\nonumber\\
&&=\frac{1}{\sqrt{2\pi}}
\frac{\sqrt{(\varepsilon-\xi_1)(\varepsilon-\xi_2)}}
{\left(\frac{\Delta^2}{3}\right)^{3/2}}
e^{-\frac{(\varepsilon-\xi_1)(\varepsilon-\xi_2)}
{2\left(\frac{\Delta^2}{3}\right)}}\times\nonumber\\
&&\times\theta[(\varepsilon-\xi_1)(\varepsilon-\xi_2)]sign(2\varepsilon-\xi_1-\xi_2)
\label{SpDenSF}
\end{eqnarray}
again with the same qualitative conclusions as in incommensurate case.

For the general case of finite correlation lengths $\xi=\kappa^{-1}$ spectral
densities can be directly computed using analytic continuation of recursion
relations (\ref{Sk}), (\ref{rec}) to real frequencies \cite{Sch,KS}.

Actually, two -- dimensional contour plots of $A(\varepsilon=0,\xi_{\bf p})$ 
(which are in direct correspondence with ARPES intensity plots) can be used
for ``practical'' definition of renormalized Fermi surface and provide a
qualitative picture of its evolution in FGM with the 
change model parameters\footnote{Note that for the free electrons 
$A(\varepsilon=0,\xi_{\bf p})=\delta(\xi_{\bf p})$, so that appropriate
intensity plot directly reproduces the ``bare''  Fermi surface.}.

\begin{figure}[htb]
\includegraphics[clip=true,width=0.5\textwidth]{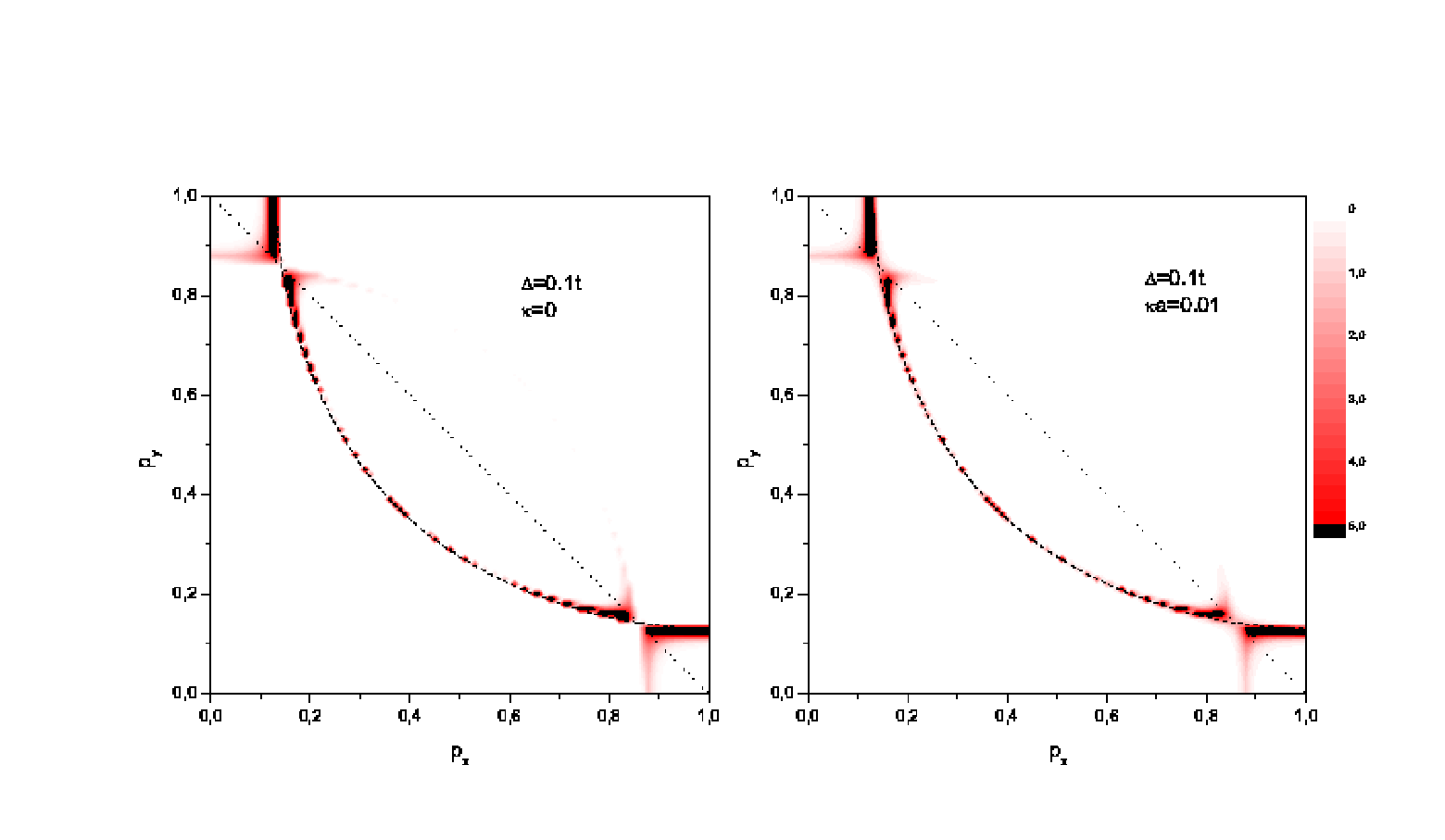}
\includegraphics[clip=true,width=0.5\textwidth]{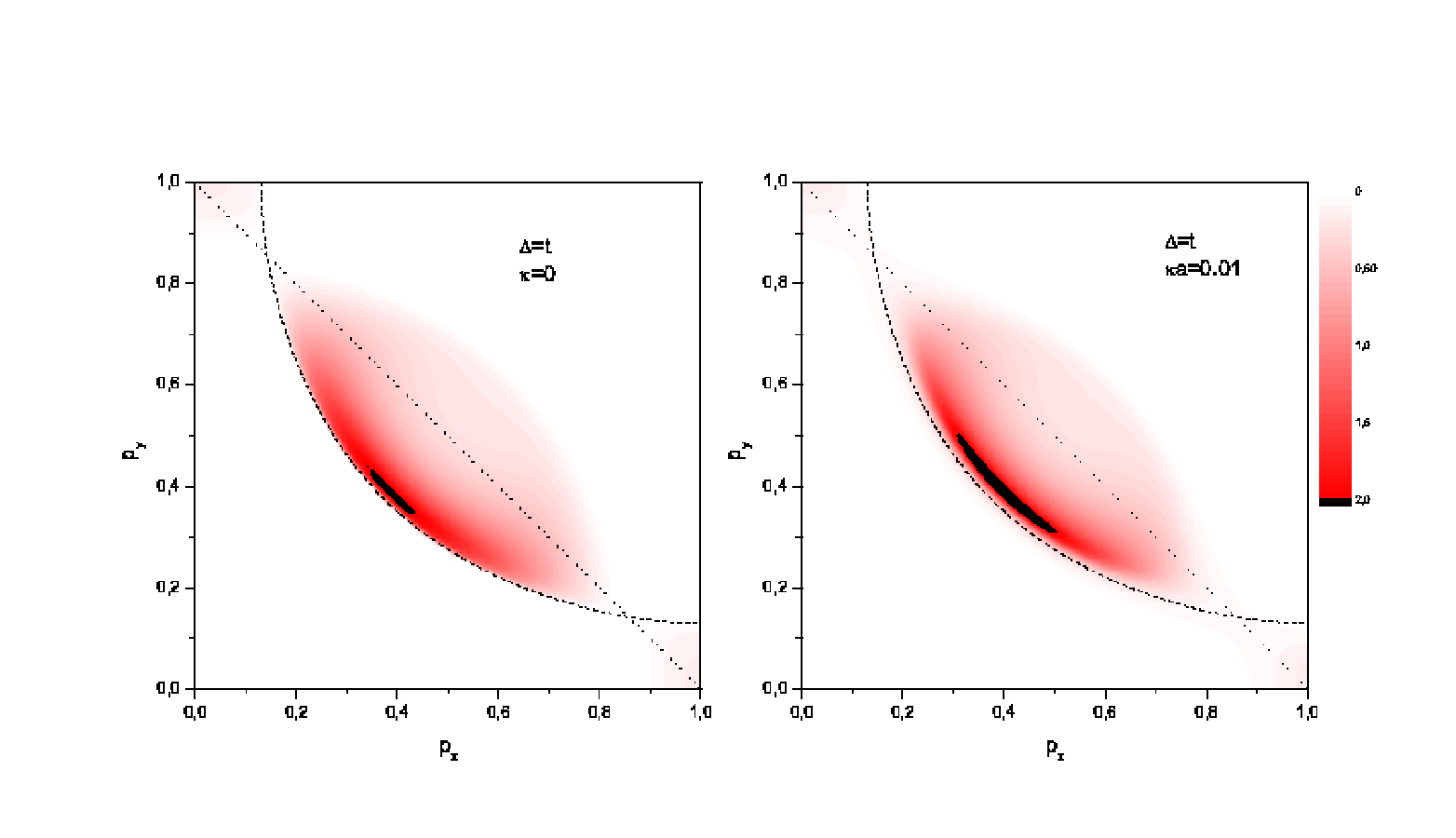}
\caption{Intensity plots of spectral density $A(\varepsilon=0,\xi_{\bf p})$ 
in Brillouin zone for the ``hot spots'' model ($t'=-0.4t$ and $\mu=-1.3t$) 
for the case of infinite correlation length $\xi^{-1}=\kappa=0$ and for 
finite correlation length $\xi^{-1}a=\kappa a= 0.01$ (spin -- fermion 
combinatorics of diagrams) with different values of pseudogap amplitude 
$\Delta$.  ``Bare'' Fermi surface is shown by dashed line.} 
\label{FS01}
\end{figure}

In Fig. \ref{FS01} we show typical intensity plots 
of spectral density $A(\varepsilon=0,\xi_{\bf p})$ in Brillouin zone for the 
``hot spots'' model both for the case of infinite correlation length 
$\xi^{-1}=\kappa=0$  and for finite (large!) correlation length 
$\xi^{-1}a=\kappa a= 0.01$ (spin -- fermion combinatorics of diagrams, 
in other cases behavior is quite similar) and for different values of 
pseudogap amplitude $\Delta$. We see that these spectral density plots give
rather beautiful qualitative picture of the ``destruction'' of the Fermi
surface in the vicinity of ``hot spots'' for small values of $\Delta$, with
formation of typical ``Fermi arcs'' as $\Delta$ grows, which is qualitatively
resembling typical ARPES data for copper oxides \cite{ARPES,Kord}.

\subsection{Superconducting $d$ -- wave fluctuations.}

As we noted above, the case of superconducting $s$ - wave pseudogap fluctuations
simply reduces to one -- dimensional FGM. Much more interesting is the case of
superconducting $d$ - wave fluctuations in $2D$.

To obtain exact results for the case of infinite correlation length 
$\xi^{-1}=\kappa=0$ we have only to make simple replacements in the above 
expressions for the ``hot spots'' model with incommensurate combinatorics:\
$\xi_2\to -\xi_1=-\xi_{\bf p}$ and $\Delta\to\Delta_{\bf p}$, where
$\Delta_{\bf p}$ defines the amplitude of fluctuations with $d$ - wave 
symmetry:
\begin{equation}
\Delta_{\bf p}=\frac{1}{2}\Delta\left[\cos (p_xa) - \cos (p_ya) \right]
\label{Delta_d}
\end{equation}
with $\Delta$ now characterizing the energy scale of pseudogap fluctuations.

Then (\ref{z_var}) reduces to $z=-(\varepsilon_n^2+\xi_{\bf p}^2)$ and for
$Z$-factor we immediately obtain an expression, similar to (\ref{Z1D}): 
\begin{eqnarray}
&&Z(\varepsilon_n, \xi_p)=\nonumber\\
&&=-\frac{\varepsilon_n^2+\xi_{\bf p}^2}{\Delta^2_{\bf p}}
\exp\left({-\frac{\varepsilon_n^2+\xi_{\bf p}^2}{\Delta^2_{\bf p}}}\right)
Ei\left(-\frac{\varepsilon_n^2+\xi_{\bf p}^2}{\Delta^2_{\bf p}}\right)\approx
\nonumber\\ 
&&\approx -\frac{\varepsilon_n^2+\xi_p^2}{\Delta^2_{\bf p}}\ln\left(\gamma' 
\frac{\varepsilon_n^2+\xi_p^2}{\Delta^2_{\bf p}}\right)\to 0\nonumber\\
&&\quad\mbox{for}\quad \varepsilon_n\to 0,\ \xi_{\bf p}\to 0 
\label{Z2D} 
\end{eqnarray}
again replacing the pole singularity by zero at the ``bare'' Fermi surface,
except the ``nodal'' at the diagonal of the Brillouin zone, where 
$\Delta_{\bf p}=0$ (cf. (\ref{Delta_d})).

Instead of (\ref{SpDenInc}), we get spectral density as:  
\begin{equation}
A(\varepsilon,\xi_{\bf p})=\frac{\varepsilon+\xi_{\bf p}}{\Delta^2_{\bf p}}
e^{-\frac{\varepsilon^2-\xi^2_{\bf p}}{\Delta^2_{\bf p}}}
\theta(\varepsilon^2-\xi^2_{\bf p})sign \varepsilon
\label{SpDenSc}
\end{equation}
which is nonzero only for $|\xi_{\bf p}|<\varepsilon$. As a result, for
$\varepsilon=0$ we have $A(\varepsilon=0,\xi_{\bf p})=0$ for 
$\Delta_{\bf p}\neq 0$, and it is different from zero only at the 
intersection of Brillouin zone diagonal with ``bare'' Fermi surface, where
$\Delta_{\bf p}$ given by (\ref{Delta_d}) is zero. 
At Fermi surface itself we have:  
\begin{equation} 
A(\varepsilon,\xi_{\bf p}=0)=\frac{|\varepsilon|}{\Delta^2_{\bf p}} 
e^{-\frac{\varepsilon^2}{\Delta^2_{\bf p}}}
\label{A_sc_FS}
\end{equation}
with two maxima at $\varepsilon=\pm \Delta/\sqrt{2}$.

\begin{figure}[htb]
\includegraphics[clip=true,width=0.5\textwidth]{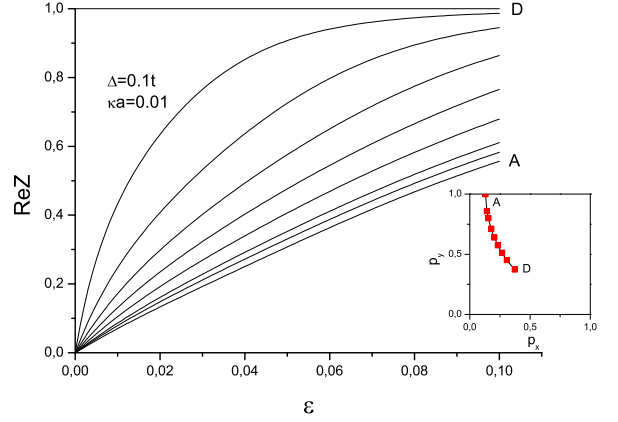}
\caption{Dependence of $Re Z$ on $\varepsilon_n$ (in units of $t$) at 
different points Fermi surface (corresponding to $t'=-0.4t$ and $\mu=-1.3t$) 
in the model of superconducting ($d$ - wave) pseudpgap fluctuations
with correlation length $\xi^{-1}a=\kappa a=0.01$. 
Pseudogap amplitude $\Delta=0.1t$. At the insert we show the 
``bare'' Fermi surface and points, where calculations were done.}
\label{Ze_SC}
\end{figure}

Considering the general case of finite correlation lengths 
$\xi=\kappa^{-1}$ we again perform numerical analysis using the recursion 
relations introduced for this problem in Ref. \cite{KS}, 
using the basic definition of $Z$ - factor given in (\ref{Z_fac}).
To calculate self -- energy $\Sigma(\varepsilon_n,\xi_{\bf p})$ 
of an electron scattered by static fluctuations of superconducting order
parameter with $d$ - wave symmetry, we use the following relation
(similar to (\ref{rec})), sligthly generalizing relations derived in
of Ref. \cite{KS}:
\begin{eqnarray} 
&&\Sigma_{k}(\varepsilon_n,\xi_{\bf p})=\nonumber\\
&&=\frac{\Delta^2_{\bf p}s(k)} {i\varepsilon_n-
(-1)^k\xi_{\bf p}+ik\kappa(|v^x_{\bf p}|+|v^y_{\bf p}|)- 
\Sigma_{k+1}(\varepsilon_n,\xi_{\bf p})} 
\nonumber\\
\label{recsc} 
\end{eqnarray} 
where $s(k)$ is defined in (\ref{vinc}),  

In Fig. \ref{Ze_SC} we show the results for 
$Re Z(\varepsilon_n, \xi_{\bf p}=0)$ again taken at different points of 
the ``bare'' Fermi surface, shown at the insert. Correlation length is
$\xi=100 a$ ($\kappa a= 0.01$) and $\Delta=0.1t$. 
It is clearly seen that $Re Z = 1$ precisely at the ``nodal'' point $D$
(where $\Delta_{\bf p}=0$), but in other point on the ``bare'' Fermi surface 
$Re Z$ is strongly renormalized in rather wide intervals of $\varepsilon_n < 
|\Delta_{\bf p}|$, going to zero with $\varepsilon_n\to 0$. Thus we again 
obtain a kind of ``marginal'' Fermi liquid or Luttinger liquid (NFL), but 
qualitatively different from the case of ``hot spots'' model.

\begin{figure}[htb]
\includegraphics[clip=true,width=0.5\textwidth]{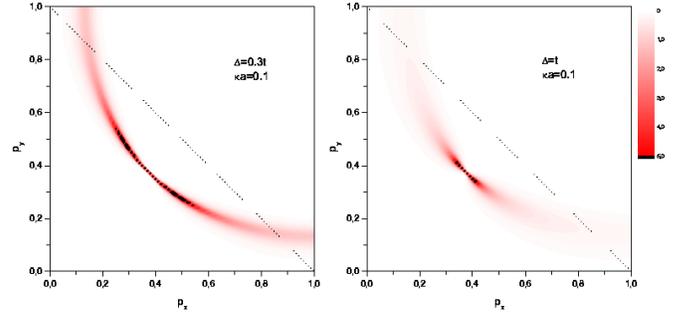}
\caption{Intensity plots of spectral density $A(\varepsilon=0,\xi_{\bf p})$ 
in Brillouin zone ($t'=-0.4t$ and $\mu=-1.3t$) 
for the case of superconducting ($d$ - wave) pseudogap fluctuations. 
Correlation length $\xi^{-1}a=\kappa a= 0.1$ (spin -- fermion 
combinatorics of diagrams) for two different values of pseudogap amplitude
$\Delta=0.3t$ and $\Delta=t$.} 
\label{FS_sc} 
\end{figure}

In Fig. \ref{FS_sc} we also show typical intensity plots 
of spectral density $A(\varepsilon=0,\xi_{\bf p})$ in Brillouin zone for the 
case of superconducting ($d$ - wave) pseudogap fluctuations with
correlation length $\xi^{-1}a=\kappa a=0.1$  and two different values of
$\Delta$. We see that these spectral density plots give quite different
picture of the ``destruction'' of the Fermi surface in comparison with
the case of ``hot spots'' model, which also, in our opinion, differs 
significantly from most results of ARPES measurements on copper oxides. 
Fermi surface is sharply defined only in one point (at the diagonal of the 
Brillouin zone), where $\Delta_{\bf p}$ given by (\ref{Delta_d}) is precisely 
zero, and there are no sharply defined Fermi arcs formed close to this point. 
We observe only some more or less wide ``dragon -- fly wings'' formed around 
this point. Note also the absence of any signs of ``shadow'' Fermi surface.

\section{Conclusion}

We analyzed rather unusual (NFL) behavior of fluctuating 
gap model (FGM) of pseudogap behavior in both $1D$ and $2D$. 
We studied in detail quasiparticle renormalization ($Z$ -- factor) of the
single -- electron Green's function, demonstrating a kind
of ``marginal'' Fermi liquid or Luttinger liquid behavior 
(i.e. the absence of well -- defined quasiparticles close to the Fermi
surface) and also the topological stability of the ``bare'' Fermi surface 
(Luttinger theorem). This reflects strong renormalization effects leading to
the replacement of the usual pole singularity of the Green's function in 
Fermi liquid by {\em zero}, thus effectively replacing at the Fermi surface
of poles by {\em Luttinger surface} of zeroes \cite{Dz}. In the
presence of static impurity -- like scattering due to the effects of finite 
correlation lengths of pseudogap fluctuations this singularity is replaced by 
finite discontinuty.

In $2D$ case we discussed effective picture of Fermi surface ``destruction'' 
both in ``hot spots'' model of dielectric (AFM, CDW) pseudogap fluctuations, 
as well as for qualitatively different case of superconducting $d$ - wave 
fluctuations, reflecting NFL spectral density behavior and similar to that
observed in ARPES experiments on copper oxides.  Intensity plots obtained for
the case of AFM (CDW) fluctuations are, in our opinion, more resembling ARPES
intensity data obtained in experiments on copper oxides. Note, that this
effective picture was also directly generalized for the case of strongly
correlated metals or doped Mott insulators \cite{KNS} using the so called
$DMFT+\Sigma_k$ approach of Ref. \cite{DMFT_S}. 

Authors are gratefull to G. E. Volovik for his interest and quite useful
discussions, which, in fact, initiated this work.

This work was supported in part by RFBR grant 05-02-16301 
and programs of the Presidium of the Russian Academy of Sciences (RAS) 
``Quantum macrophysics'' and of the Division of Physical Sciences of the RAS 
``Strongly correlated electrons in semiconductors, metals, superconductors and 
magnetic materials''.



\end{document}